\documentclass[referee]{raa}            

\usepackage{graphicx,times}             
\usepackage{natbib}
\bibpunct{(}{)}{;}{a}{}{,}
\usepackage{cuted}
\usepackage{graphicx}	
\usepackage{amsmath}	
\usepackage{amssymb}	
\usepackage{amsfonts}
\usepackage{ulem}
\usepackage{pdfpages}
\usepackage[dvipsnames]{xcolor}  
\usepackage{soul}

\usepackage[pagebackref=true]{hyperref}
\DeclareUnicodeCharacter{2212}{-}
\DeclareUnicodeCharacter{00A0}{~}
\begin{document}

  \title{Broadband Spectral properties of MAXI J1348--630 using {\textit{AstroSat}} Observations}

   \volnopage{Vol.0 (20xx) No.0, 000--000}      
   \setcounter{page}{1}          

   \author{Gitika Mall 
      \inst{1}
   \and Jithesh Vadakkumthani
      \inst{2}
   \and Ranjeev Misra
      \inst{3}
   }

   \institute{Center for Field Theory and Particle Physics and Department of Physics, Fudan University, Shanghai-200438, China; {\it gitikamall20@fudan.edu.cn}\\
        \and
             Department of Physics and Electronics, CHRIST (Deemed to be University), Hosur Main Road, Bengaluru - 560029, India\\
        \and
             Inter-University Centre for Astronomy and Astrophysics (IUCAA), PB No.4, Ganeshkhind, Pune-411007, India\\
\vs\no
   {\small Received 20xx month day; accepted 20xx month day}}

\abstract{We present broadband X-ray spectral analysis of the black hole X-ray binary MAXI J1348--630, performed using five {\it AstroSat} observations. The source was in the soft spectral state for the first three and in the hard state for the last two observations. The three soft state spectra were modelled using a relativistic thin accretion disc with reflection features and thermal Comptonization. Joint fitting of the soft state spectra constrained the spin parameter of the black hole  $a_*$ $>$ 0.97 and the disc inclination angle $i$ = 32.9$\substack{+4.1 \\ -0.6}$ degrees. The bright and faint hard states had bolometric flux a factor of $\sim 6$ and $\sim 10$ less than that of the soft state. Their spectra were fitted using the same model except that the inner disc radius was not assumed to be at the last stable orbit. However, the estimated values do not indicate large truncation radii and the inferred accretion rate in the disc was an order of magnitude lower than that of the soft state. Along with earlier reported temporal analysis, {\it AstroSat} data provides a comprehensive picture of the evolution of the source.
}

   \authorrunning{Mall et al.}            
   \titlerunning{ Broadband Spectral properties of MAXI J1348--630 }  

   \maketitle
%
%
\section{Introduction}           
\label{sec:intro}
Transient X-ray binaries, particularly black hole X-ray binaries (BHXRBs), spend long periods in quiescence and occasionally exhibit outbursts typically lasting from weeks to a few months. During the outburst, transitions between various X-ray spectral states accompanied by variations in spectral properties are observed (\citealt{Homan2005}). These states, characterised by luminosities and spectral shapes, are primarily of two types: Low Hard State (LHS) and High Soft State (HSS). There are also intermediate states. Typically, in the rising phase of the outburst, the source evolves from an LHS to Hard-intermediate State (HIMS) followed by Soft-intermediate State (SIMS) and finally to an HSS (\citealt{Homan2005}; \citealt{Remillard_2006}; \citealt{Belloni_2009}).  During the decaying phase of the outburst, the source returns to LHS from HSS via intermediate state (\citealt{Motta_2011}; \citealt{Homan2005}; \citealt{Mu_oz_Darias_2011}; \citealt{2006csxs.book...39V}).\\
In the soft state, the X-ray flux is dominated by thermal emission from an optically thick and geometrically thin accretion disc (the standard disc; \citealt{1973A&A....24..337S}). This observed spectrum can be reproduced with a multi-colour disc model (\citealt{1984PASJ...36..741M}). The inner disc radius remains constant in this state irrespective of significant changes in X-ray luminosity and temperature (e.g., \citealt{1993ApJ...403..684E}; \citealt{2004ApJ...601..428K}; \citealt{Steiner_2010}; \citealt{10.1093/pasj/63.sp3.S803}). This implies that the standard disc extends down to the innermost stable circular orbit (ISCO). Since the inner edge of the disc is close to the black hole, the observed disc black body spectra can be fundamentally altered by Doppler shifting and relativistic effects (\citealt{Gierlinski_2004}). The resulting spectral shape depends strongly on the viewing angle to the source (\citealt{8185179}). In this state, the disc spectrum is prominent enough to measure the spin of the compact object using the continuum-fitting method. This is done by estimating the inner radius (R\textsubscript{in}) of the accretion disc and identifying it with the radius of ISCO (R\textsubscript{ISCO}). R\textsubscript{ISCO} is a monotonic function of the dimensionless black hole spin parameter $a_*$, decreasing from 6 r$_{g}$ to 1.235 r$_{g}$ (where r$_{g}$ is gravitational radius in units of $GM/c^2$)
as spin increases from $a_*$ = 0 to $a_*$ = 1 \citep{1972ApJ...178..347B}. This relationship between $a$ and R\textsubscript{ISCO} allows for estimating the spin parameter by continuum-fitting method (\citealt{Zhang_1997}; \citealt{Gou_2011}). A complementary method for spin measurement is to use the shape of the relativistically modified fluorescence iron line emission, which depends on the disc inclination angle (\citealt{1989MNRAS.238..729F}; \citealt{2003PhR...377..389R}).\\
In the hard state, the energy spectra are dominated by a strong Comptonized emission from the hot cloud of electrons known as the ``corona'' (\citealt{2008MmSAI..79..118P}) and the source exhibits a hard power-law shaped X-ray spectrum with an index $\sim 1.7$. Moreover, the disc appears much cooler compared to the soft state and is believed to be truncated at a large radius from the compact object \citep{Remillard_2006}. The quasi-steady jets are often associated with this state and clear correlations between the radio and X-ray intensities are observed  (\citealt{Remillard_2006}; \citealt{Belloni_2009}). However, contrary to the truncated disc model, in a few cases, the disc has been found to extend close to the black hole in the hard state, when the source luminosity is greater than a few per cent of Eddington luminosity (\citealt{Shidatsu_2014}). For example, modelling the reflection spectra of GX 339--4 in two extreme spectral states, \citet{10.1111/j.1365-2966.2008.13358.x} reported an inner radius of r\textsubscript{in}= 2.08$\substack{+0.17 \\ -0.10}$ r\textsubscript{g}.\\
The new X-ray transient source MAXI J1348--630 was discovered by the {\it MAXI/GSC} instrument on 2019 January 26 (\citealt{2019ATel12425....1Y}). The reported position of the source is R.A. = 13:48:12.73, Decl. = -63:16:26.8 (equinox J2000.0) with an uncertainty of $\sim$ 1.7 arcsec (90 confidence level; \citealt{2019ATel12434....1K}) using the {\it Swift} XRT observations. After this discovery, all major X-ray observatories such as {\it INTEGRAL}, {\it NICER} and {\it Insight-HXMT} (\citealt{2019ATel12441....1A}; \citealt{2019ATel12447....1S}; \citealt{2019ATel12470....1C}) conducted follow-up observations to understand the nature of the source. Optical and radio counterparts (\citealt{2019ATel12430....1D}; \citealt{2019ATel12456....1R}) have been reported for the source and the multi-wavelength properties suggested that the source is a black hole X-ray binary (\citealt{2019ATel12456....1R}; \citealt{Carotenuto_2021}). {\it NICER}  monitoring observations showed outburst evolution similar to what is commonly observed in BHXRBs and during the outburst phase, the source exhibited different types of quasi-periodic oscillations as well(\citealt{2020MNRAS.499..851Z}; \citealt{2020MNRAS.496.4366B}). \citet{Tominaga_2020} reported the analysis of the first half-year {\it MAXI} monitoring of the source. In these observations, the source exhibited spectral transitions between the soft and hard states and was consistent with the well-known characteristic of a BHXRB. Assuming a face-on disc around a non-spinning black hole, they estimated the source distance and the black hole mass to be $\approx$ 4 kpc and $\sim$ 7 (D/4 kpc) M$_\odot$, respectively and noted that the black hole could be more massive depending on the disc inclination and the spin parameter. Modelling the combined {\it Swift} XRT, BAT, and {\it MAXI/GSC} spectra in the 1--150 keV energy band using the two-component advective flow (TCAF) model, \citet{Jana_2020} estimated the black hole mass $\sim$ 9 M$_\odot$ and the source distance to be 5--10 kpc. \citet{Chauhan_2020} studied the HI absorption spectra and obtained a probable distance of {2.2$\substack{+0.5 \\ -0.6}$} kpc with a strong upper limit of $5.3 \pm 0.1$ kpc. Later, \citet{Lamer_2021} combined data from multiple satellites and used the dust scattering halo to precisely measure the source distance to be 3.39 kpc (with a statistical uncertainty of 1.1\%) and a renewed mass estimate for the compact object at $11\pm2$ M$_\odot$. Using the relativistic reflection model, \citet{Jia_2022} fitted the {\it NuSTAR} spectra of the source and obtained the the spin parameter $a_{*}$ = {0.78$\substack{+0.4\\-0.4}$}, and the inclination angle of the inner disc $i$ = {29.2$\substack{+0.3\\-0.5}$} degrees. Recently, \citet{2022MNRAS.513.4869K} studied the source using the near-simultaneous {\it NICER} and {\it NuSTAR} observations. They estimated the spin, black hole mass and inclination of the source by fixing the distance to 2.2 kpc, and the values were found to be $0.80 \pm 0.02$, $8.7 \pm 0.3$ M$_\odot$ and $36.5 \pm 1.0$ degrees, respectively.
\citet{Jithesh_2021} used five simultaneous {\it AstroSat} and {\it NICER} observations for studying the broadband X-ray spectral-timing properties of the source. They noted the presence of quasi-periodic oscillations of type-C and type-A classes and further via power density spectrum and energy-dependent rms, they confirmed that the disc is significantly less variable than the Comptonization component. They detected hard time lag for the bright hard state in the 0.5--80 keV energy band and modelled the energy-dependent rms and time lag using a single-zone propagation model. They used a multi-colour disc emission, Gaussian line and thermal Comptonization to identify the spectral states of the observations.\\
In this work, we study the broadband X-ray spectral characteristics of MAXI J1348--630 using the same {\it AstroSat} observations as used by \citet{Jithesh_2021}. The primary aim is to constrain the system parameters such as the spin parameter and the inclination angle of the accretion disc using relativistic spectral models. We have utilized data from the LAXPC and SXT payloads onboard {\it AstroSat}. § \ref{sec:AstroSat} describes the observations used in this work and the data reduction techniques. The broadband spectral analysis methodology is put forward in §\ref{sec:3}. The main results are discussed in §\ref{sec:4}.

\section{Observations and Data Reduction}
\label{sec:AstroSat} 
{\it AstroSat} (\citealt{2014SPIE.9144E..1SS}; \citealt{Agrawal_2017}) observed MAXI J1348--630 during its 2019 outburst and we avail five of such observations for our study. The details of observations used in this work are given in Table \ref{tab:table1}.

\begin{table*}
	\caption{Observation Log. (1) Data; (2) Observation ID; (3) date of observation; (4) exposure time (L and S represent LAXPC and SXT respectively); (5) LAXPC count rate in the energy range of 4--25 keV; (6)  SXT count rate in the 0.8--5 keV energy band; (7) inner and outer radius of the annulus used for extraction of SXT spectrum. $\textsuperscript{a}$Since the SXT data is not piled-up in Data4 observation, we extracted the SXT spectrum from a circular region of radius 16 arcmin.}
\setlength{\tabcolsep}{3.5pt}
\small
	\begin{tabular}{l c c c c c r}
	    \hline
		\hline
		Data & ObsID & Date & Exposure & LAXPC & SXT & Radius\\
		& & & (ks) & (counts s$^{-1}$) & (counts s$^{-1}$) & (arcmin)\\
		\hline
        Data1 & T03\textunderscore 083T01\textunderscore9000002722 & 2019 February 19–20 & 5.5(L) / 1.9(S) & 1535 & 837 & 8 \& 16\\
		Data2 & T03\textunderscore 083T01\textunderscore9000002728 & 2019 February 22 & 20.2(L) / 11.1(S) & 1492 & 802 & 8 \& 16\\
		Data3 & T03\textunderscore 083T01\textunderscore9000002742 & 2019 February 28 & 23.2(L) / 12.2(S) & 1130 & 729 & 6 \& 16\\
		Data4 & T03\textunderscore 112T01\textunderscore 9000002896 & 2019 May 8–9 & 13.8(L) / 6.8(S) & 112 & 13 & 16$\textsuperscript{a}$\\
		Data5 & T03\textunderscore 120T01\textunderscore 9000002990 & 2019 June 14–15 & 35.0(L) / 14.9(S) & 922 & 65 & 2 \& 16\\
		\hline
	\end{tabular}
	\label{tab:table1}
 \end{table*}

\subsection{Large Area X-ray Proportional Counter (LAXPC)}
LAXPC has three identical proportional counters, namely LAXPC10, LAXPC20, and LAXPC30, functioning in the broad energy band of 3--80 keV with an excellent time resolution of 10 $\mu$s (\citealt{Yadav_2016}; \citealt{Antia_2017}; \citealt{Agrawal_2017}). LAXPC software (LaxpcSoft; version as of 2020 August 4) was used to process the Level-1 Event Analysis (EA) mode data. The data reduction and the extraction of science products were carried out using standard tools available in LaxpcSoft. Among the three detectors, LAXPC30 was switched off in 2018 March due to the gas leakage and LAXPC10 was operating at low gain. Thus, we used only the LAXPC20 detector in this study and modelled the spectrum in the 4-25 keV energy band.

\subsection{Soft X-ray Telescope (SXT)}
SXT is an imaging telescope and sensitive to the soft X-ray band, particularly in the 0.3--8 keV (\citealt{2016SPIE.9905E..1ES}; \citealt{2017JApA...38...29S}). The SXT observations used in this work are taken in the Photon Counting (PC) mode. Using the SXT pipeline software (version: AS1SXTLevel2-1.4b), we processed the Level-1 data obtained, cleaned the Level-2 event files from different orbits and merged them employing the SXT event merger tool\footnote{\url{http://www.tifr.res.in/~astrosat\_sxt/dataanalysis.html}}. Four out of five (except Data4) observations had pile-up, which was removed by extracting source events from an annulus region. The extraction radius used for different observations are given in Table \ref{tab:table1}. We used the blank sky SXT spectrum as the background spectrum and the ``sxt\_pc\_mat\_g0to12.rmf'' file as the redistribution matrix file (RMF). We generated the off-axis auxiliary response files (ARF) using the {\tt sxtARFModule} tool by giving the  on-axis ARF (sxt\_pc\_excl00\_v04\_20190608.arf) as input. The individual energy spectra of all these observations were re-binned logarithmically using {\tt grppha}. When modelling the spectra, the gain of the response file is modified by using the {\tt gain fit} command in {\sc XSPEC}. The slope at unity was fixed leaving the offset as a free parameter. We modelled the SXT spectrum in the 0.8--5 keV energy range. 

\section{BROADBAND X-RAY SPECTRAL ANALYSIS}
\label{sec:3}
As mentioned in § \ref{sec:intro}, \citet{Jithesh_2021} characterized the spectral states of MAXI J1348--630 using these same five {\it AstroSat} observations. They identified that the first three observations (Data1, Data2, and Data3) were in the soft spectral state of BHXRBs, while the last two observations were in the hard spectral state (Data4 and Data5). In the following two sections, we separately describe the details of the spectral modelling performed on the source in soft and hard states, respectively. 

\subsection{Soft State Observations}
\label{sec:3.1}
 
\begin{figure*}
\centering
	{\includegraphics[width=2.9in,angle=0,trim=0 0 0 0,clip]{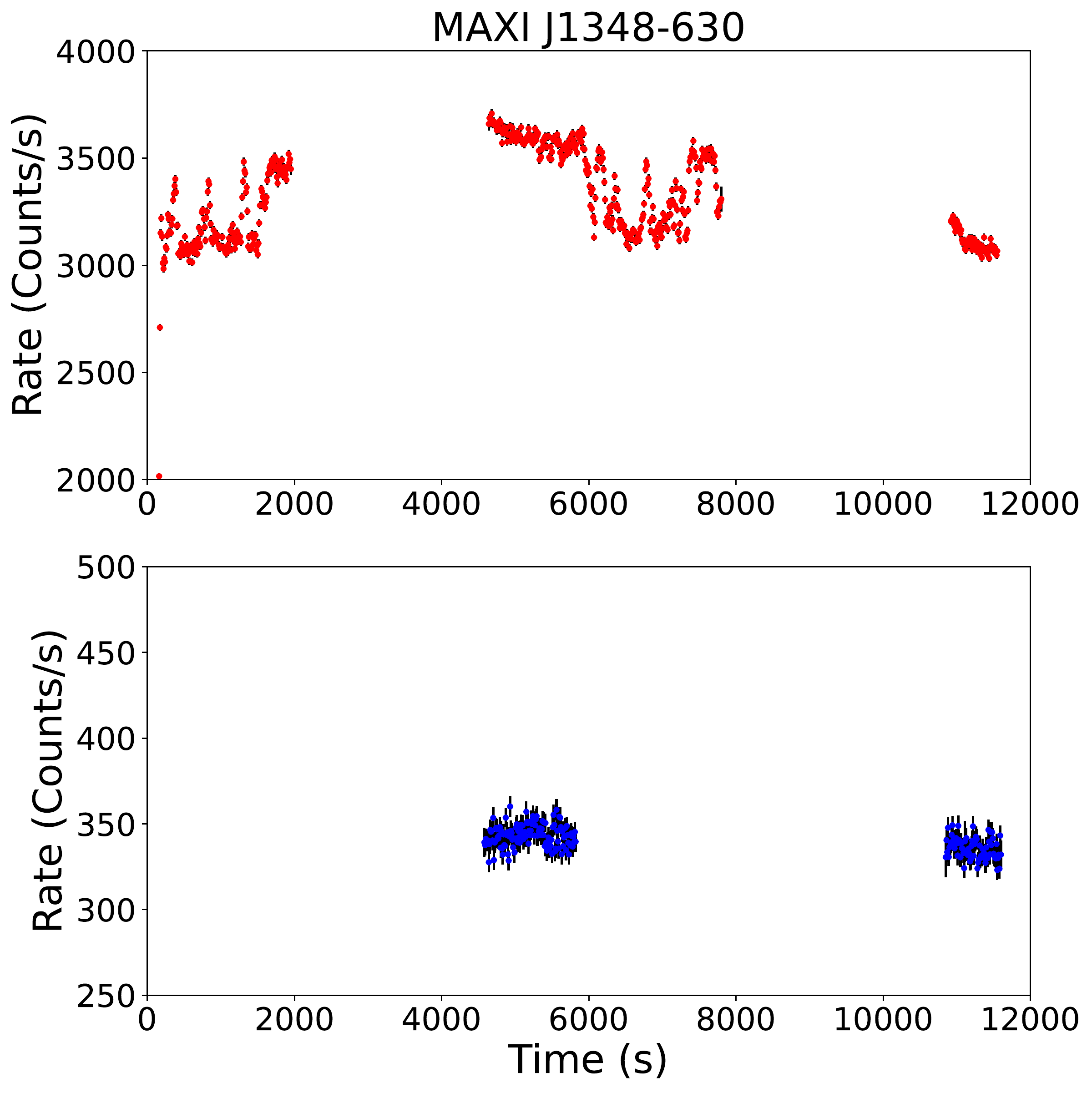}}
	{\includegraphics[width=2.9in,angle=0,trim=0 0 0 0,clip]{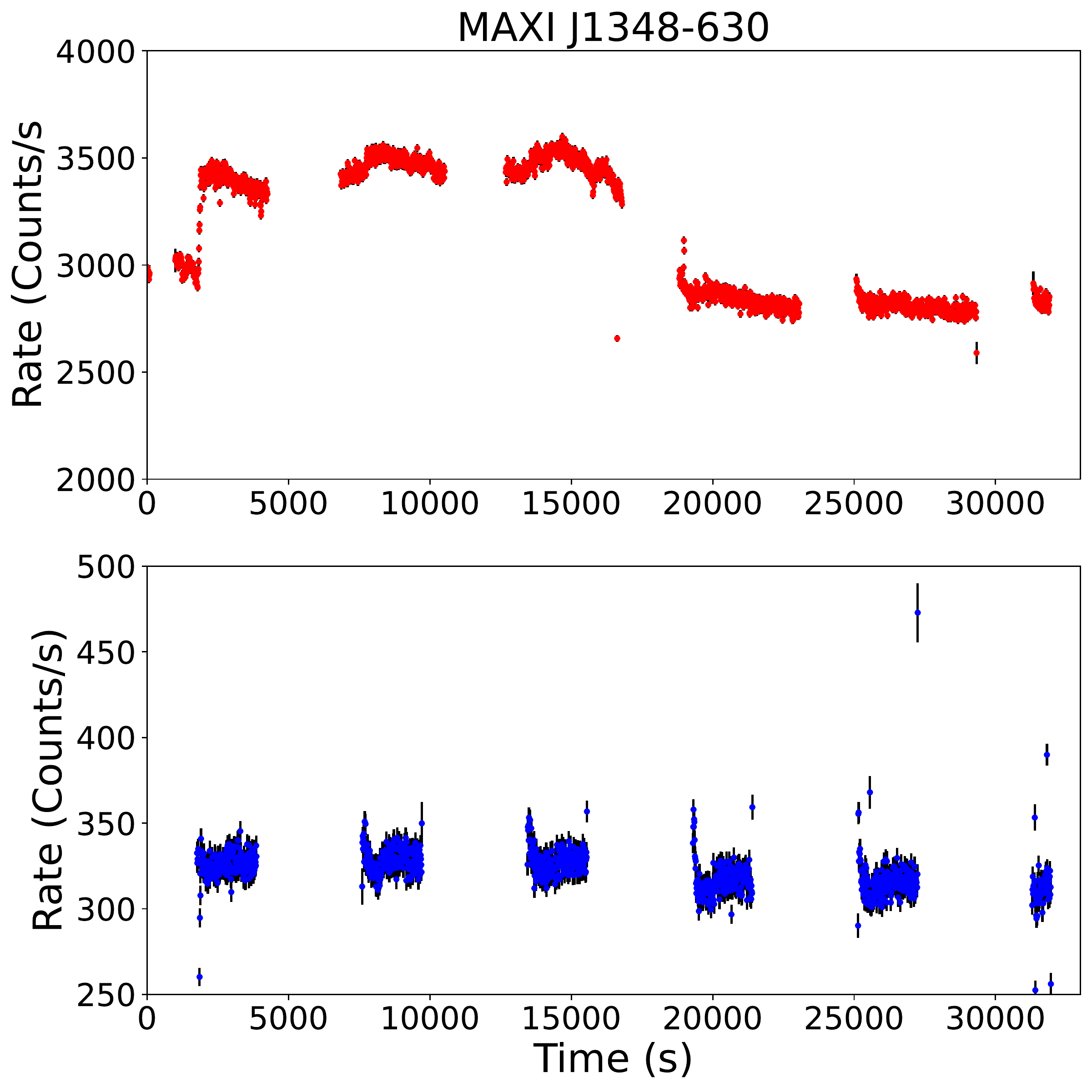}}
	{\includegraphics[width=2.9in,angle=0,trim=0 0 0 0,clip]{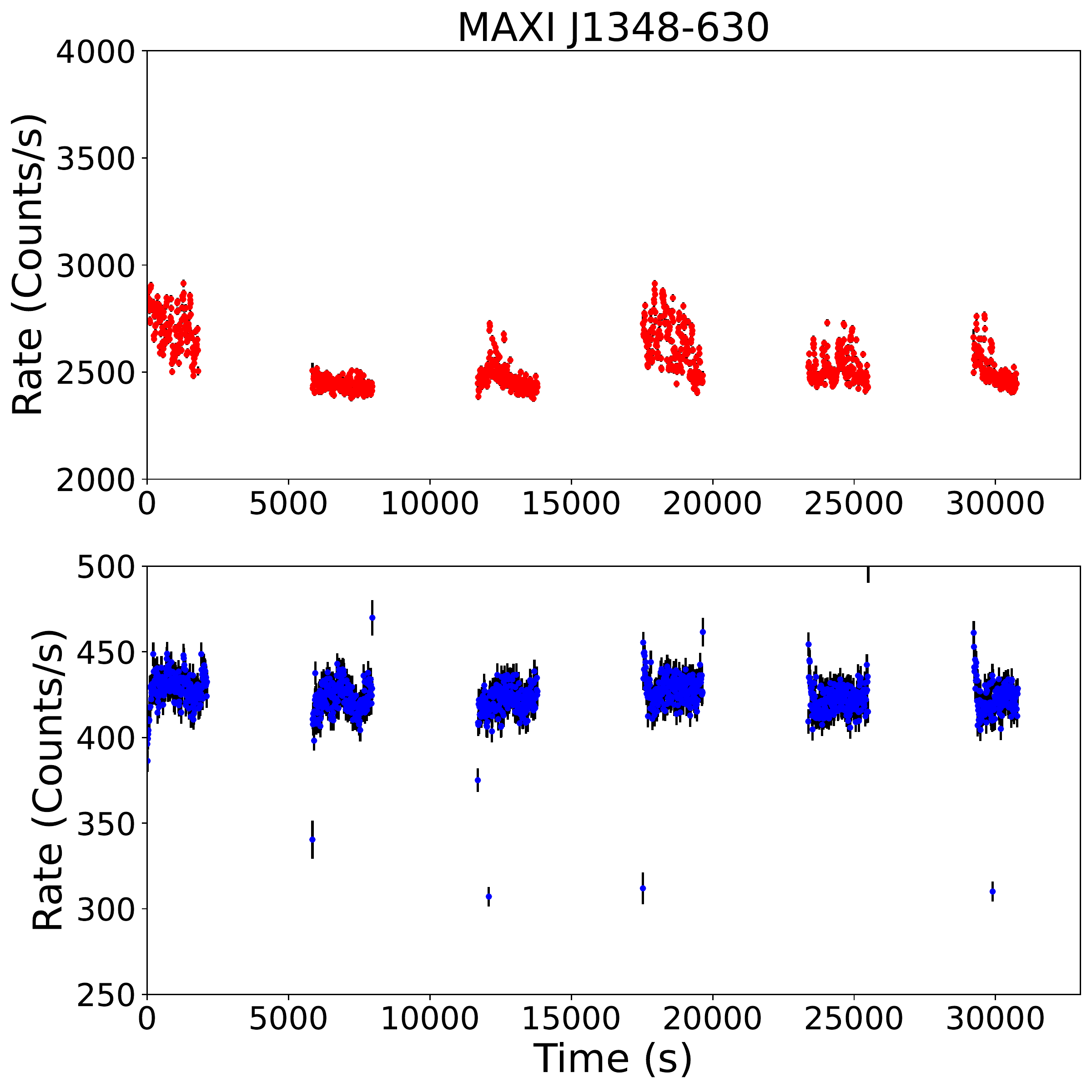}}
    \caption{The light curves from {\it AstroSat} observations in the soft state: Data1 (left), Data2 (right) and Data3 (bottom). In each plot, the top and bottom panels represent the 3--80 keV light curve from LAXPC20 binned with 10-s and the 0.3--8 keV light curve from SXT, respectively.  The reference time for Data 1 is MJD 58533, for Data 2 is MJD 58536 and, Data 3 is MJD 58542.}
    \label{fig:figure1}
\end{figure*}

\begin{figure*}
\centering
{\includegraphics[width=2.1in,angle=270,trim=0 0 0 0,clip]{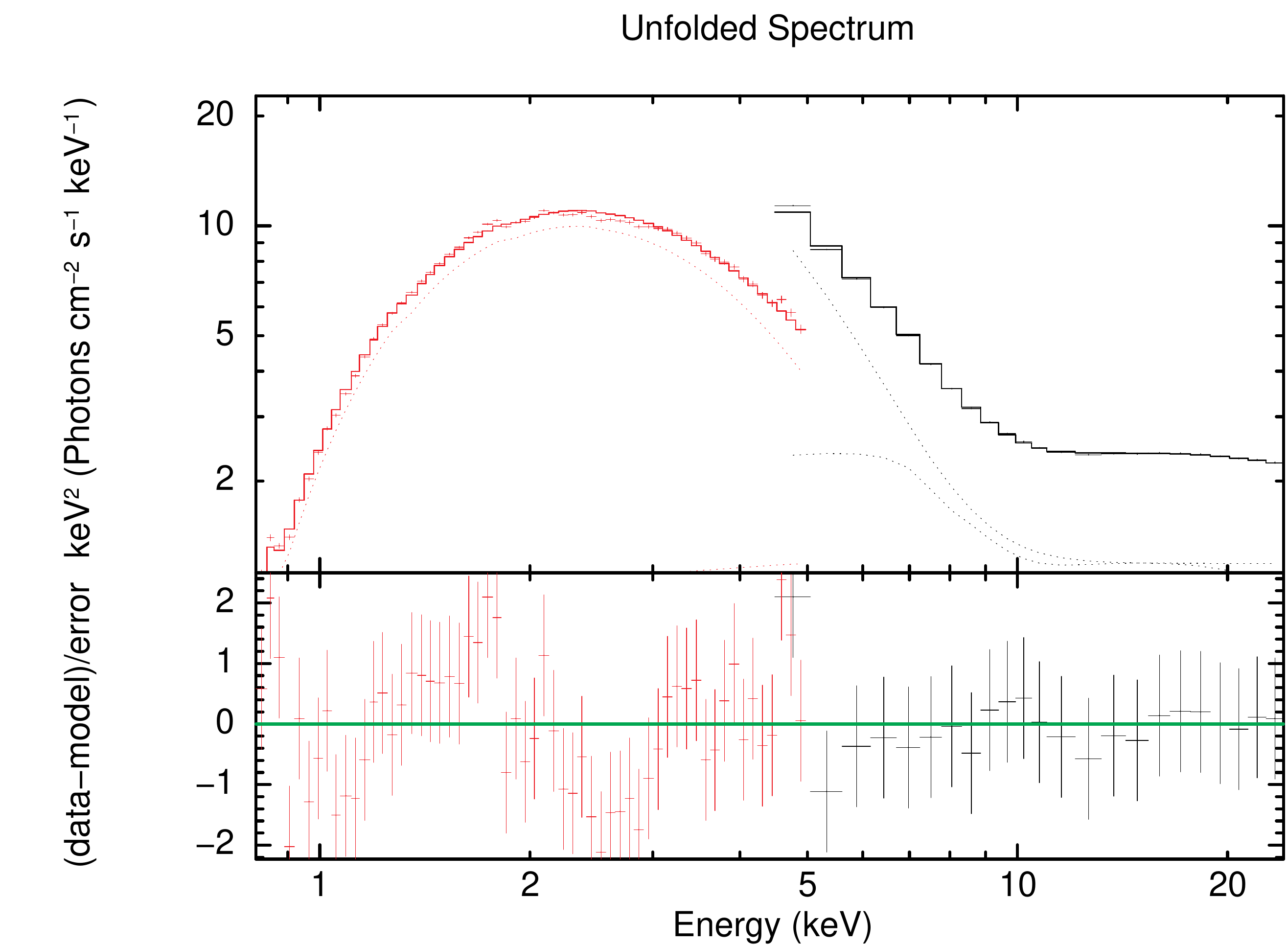}}
{\includegraphics[width=2.1in,angle=270,trim=0 0 0 0,clip]{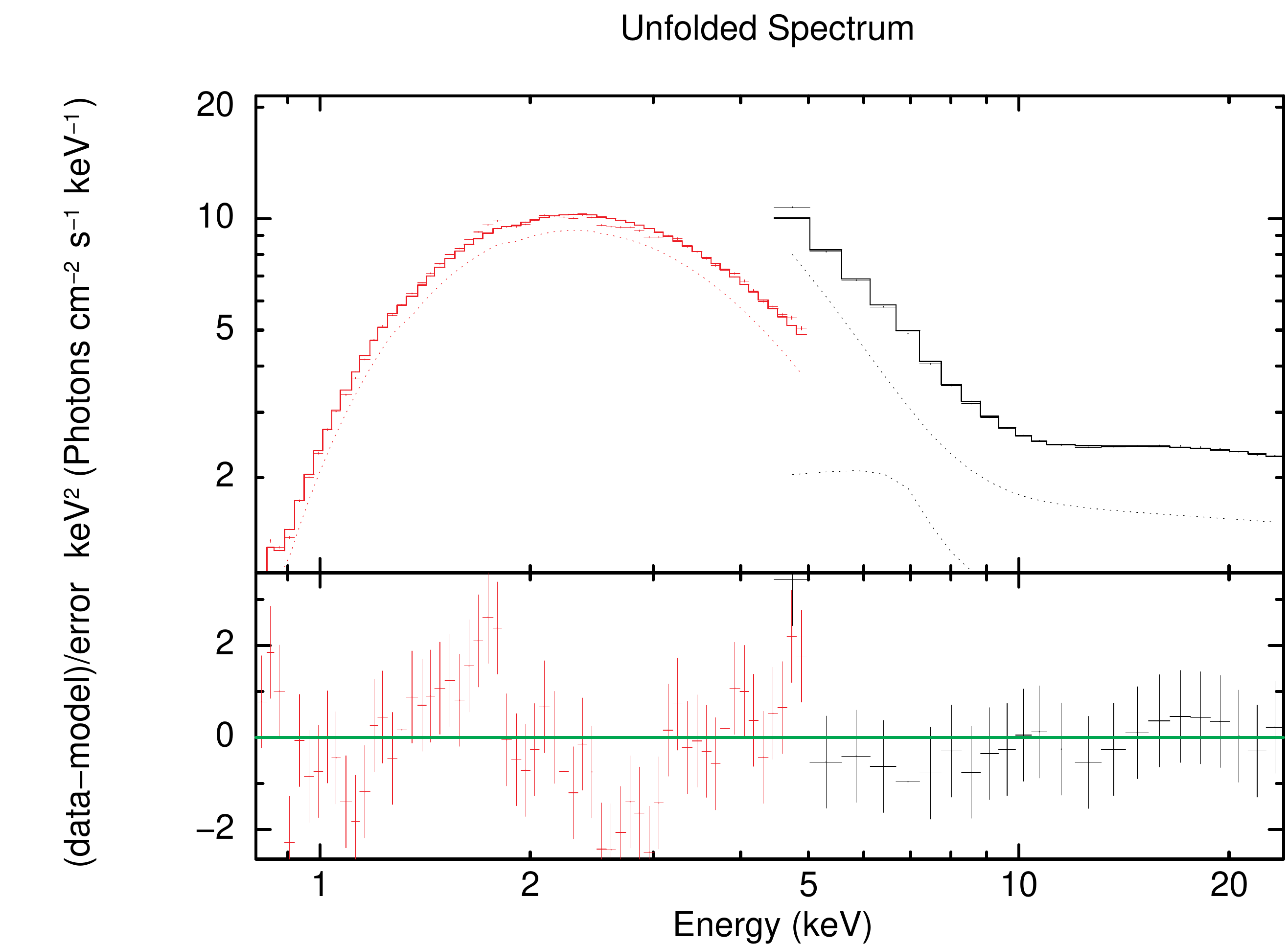}}	{\includegraphics[width=2.1in,angle=270,trim=0 0 0 0,clip]{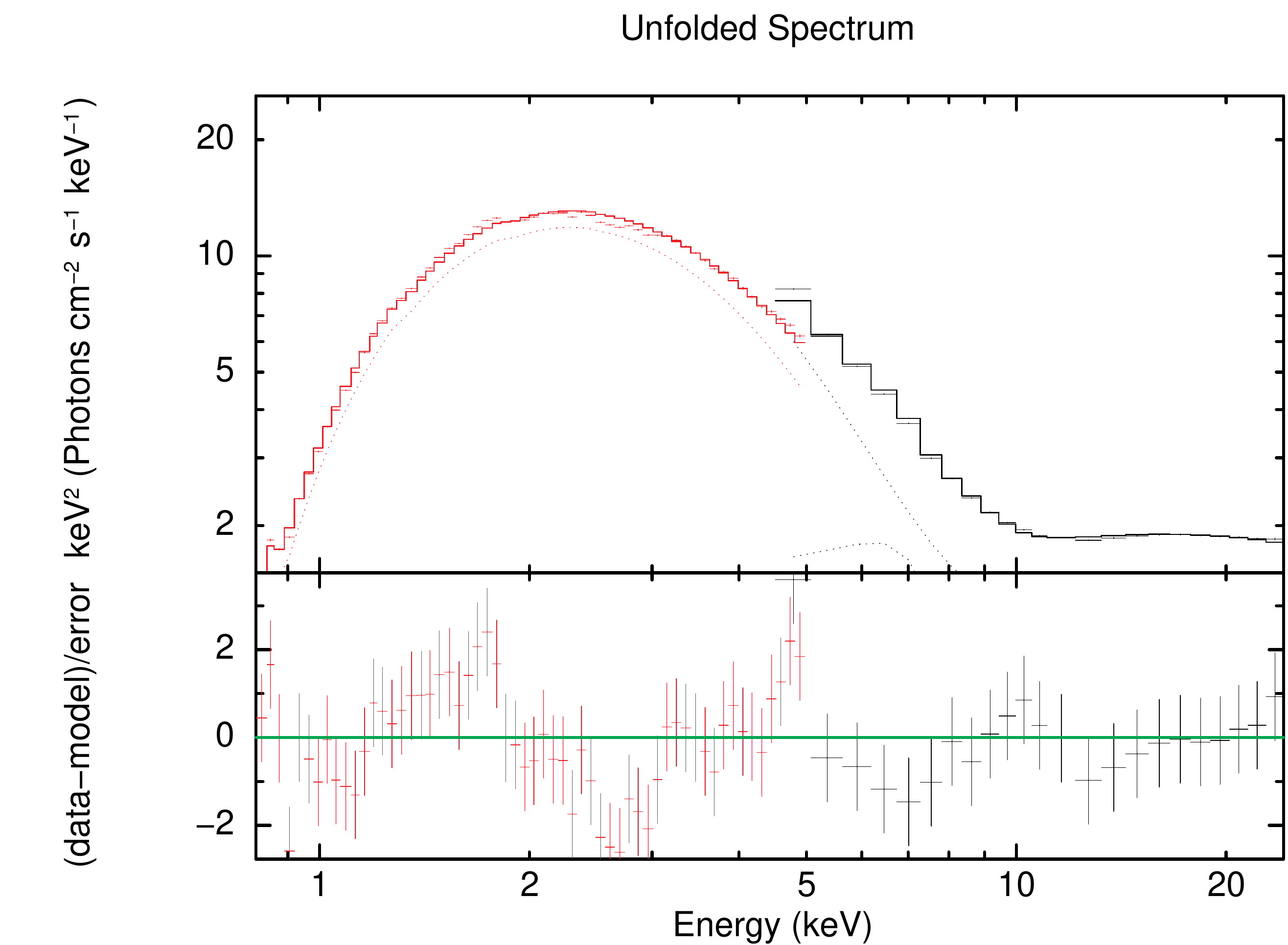}}
    \caption{The 0.8–25 keV broadband X-ray  unfolded spectra of MAXI J1348--630 from Data1 (left), Data2 (right), Data3 (bottom) observations. The black and red data points represent LAXPC20 and SXT data, respectively. Spectra are fitted with {\tt tbabs} * ({\tt simpl} * {\tt kerrbb} + {\tt relxill}).}
    \label{fig:figure2}
\end{figure*}

The background-subtracted 3--80 keV LAXPC and 0.3--8 keV SXT light curves of MAXI J1348--630 during it's soft state (Data1, Data2 and Data3) are depicted in Figure \ref{fig:figure2}. We performed broadband X-ray spectral analysis using 0.8--5 keV SXT and 4--25 keV LAXPC spectra of each observation. The spectra were analysed simultaneously by making use of the X-ray spectral fitting software {\sc xspec} version 12.11.1 (\citealt{1996ASPC..101...17A}). Here our aim was to constrain the system parameters: spin ($a_{*}$), the inclination angle of the disc ($i$), and accretion rate (${\dot{m}}$). We first employed a phenomenological model consisting of a multi-colour disc blackbody ({\tt diskbb}; \citealt{1984PASJ...36..741M}), a convolution model ({\tt simpl}; \citealt{2009PASP..121.1279S}) to represent the Comptonization of disc photons in the inner flow and a Gaussian line profile ({\tt gaussian}). We used the Tuebingen-Boulder Inter-Stellar Medium absorption model ({\tt tbabs}) to account for the Galactic absorption (\citealt{2000ApJ...542..914W}). A multiplicative constant was used to address the cross-calibration uncertainties between the SXT and LAXPC instruments. Due to uncertainties in the calibration, 2\% model systematic uncertainty was exercised for spectral modelling. The line energy of the Gaussian was fixed at 6.4 keV. The best fit parameters with the errors at 90\% confidence level are listed in Table \ref{tab:table2}.

\begin{table*}
\bc
\caption{Broadband X-ray spectral parameters for MAXI J1348--630 from soft state observations fitted with {\tt tbabs} * ({\tt simpl} * {\tt diskbb} + {\tt gaussian}) model. (1) Data; (2) neutral hydrogen column density in units of $10^{22}~\rm cm^{-2}$; (3) asymptotic power-law index; (4) scattered fraction of seed photons; (5) black hole spin parameter; (6) inner disk temperature in keV; (7) line width in keV; (8) gaussian normalization in photons cm$^{-2}$ $s^{-1}$ (9) bolometric luminosity in the units of $\rm 10^{37}~erg~s^{-1}$; (10) statistics and degrees of freedom.}
\setlength{\tabcolsep}{3.5pt}
\small
 \begin{tabular}{ccccccccccccc}
  \hline
  \hline
Data & N$_{H}$ & $\Gamma$ & $\rm F_{Sca}$ (\%) & kT$_{in}$ & N$_{disk}$ & $\sigma$ & N$_{gauss}$ & L$_{B}$ & $\chi^2/d.o.f$\\ \hline 
{Data1} & {0.53$\substack{+0.01 \\ -0.01}$} &    {2.01$\substack{+0.06 \\ -0.06}$} &    {4.9$\substack{+0.5 \\ -0.5}$} &    {0.82$\substack{+0.01 \\ -0.01}$} &    {9123$\substack{+713 \\ -663}$} &    {1.21$\substack{+0.29 \\ -0.27}$} &    {0.061$\substack{+0.017 \\ -0.017}$} &    {6.26$\substack{+0.27 \\ -0.28}$} &    {85.36/73} \\\hline 
{Data2} & {0.49$\substack{+0.01 \\ -0.01}$} &    {2.11$\substack{+0.06 \\ -0.06} $} &    {5.6$\substack{+0.6 \\ -0.6}$} &    {0.82$\substack{+0.01 \\ -0.01}$} &    {8514$\substack{+553 \\ -589}$} &    {1.16$\substack{+0.28 \\ -0.26}$}  &    {0.059$\substack{+0.017 \\ -0.017}$} &    {5.04$\substack{+0.19 \\ -0.20}$} &    {120.48/73} \\\hline
{Data3} & {0.50$\substack{+0.01 \\ -0.01}$} &    {2.04$\substack{+0.06 \\ -0.06} $} &    {4.8$\substack{+0.5 \\ -0.4}$} &    {0.81$\substack{+0.01 \\ -0.01}$} &    {7388$\substack{+505 \\ -474}$} &    {1.19$\substack{+0.23 \\ -0.25}$} &    {0.049$\substack{+0.013 \\ -0.013}$} &    {4.79$\substack{+0.05 \\ -0.17}$} & {131.07/73} \\\hline
	\end{tabular}
	\label{tab:table2}
	\ec
 \end{table*}

 For taking into account relativistic effects and reflection features, we used a model which incorporates a multi-temperature black body model for a thin accretion disc around a Kerr black hole ({\tt kerrbb}; \citealt{2005ApJS..157..335L}),and the relativistic reflection model ({\tt relxill}; \citealt{Garc_a_2014}; \citealt{Dauser_2014}) along with the convolution model ({\tt simpl}). The photon index $\Gamma$ of {\tt relxill} was tied to that of {\tt simpl}. The black hole spin parameter and inclination angle of {\tt relxill} was tied to that of {\tt kerrbb}. The spectral hardening factor and the {\tt kerrbb} normalization were fixed at 1.7 (\citealt{1995ApJ...445..780S}) and unity, respectively. For {relxill}, the emissivity profile is employed as broken power law and the emissivity index of the inner and outer part of the accretion disc was frozen at 3. The geometry of the accretion disk is described by two parameters: the inner edge of the disk which we assumed at the innermost stable circular orbit and the outer edge of the disk which was frozen to its default value of 400~r$_{\rm g}$. The redshift of the source was set to zero. The high-energy cutoff was frozen at 70 keV. The material of the disk, also characterized by the iron abundance, A$_{\rm Fe}$, is measured in units of iron Solar abundance. The ionization of the accretion disk is described by the ionization parameter $\xi$. Lastly, since the output of this model is the reflection spectrum of the accretion disk and the power law component from the corona, there are two parameters to regulate the normalizations of these two components: the normalization of the model and the reflection fraction, which regulates the relative intensity between the reflection spectrum and the power law spectrum. It is frozen to $-1$, thus the output of these models is only the reflection component.
Black hole mass and the distance to the source were fixed at $11 M_\odot$ and 3.39 kpc, respectively, as estimated by \citet{Lamer_2021}.

The unfolded spectra and residuals from the three soft state observations fitted using {\tt tbabs} * ({\tt simpl} * {\tt kerrbb} + {\tt relxill}) are depicted in Figure \ref{fig:figure2} and the best-fit model parameters are listed in Table \ref{tab:table3}. We note that the spectral fit parameters do not differ significantly between the observations. In particular, the absorption column density and the photon index seem to be constant at $\sim 6 \times 10^{21}~\rm cm^{-2}$ and $\sim$ 2.0, respectively. The estimated value of black hole spin is constrained to be $>$ 0.906 in these observations. The inferred value of inclination angle $i$ varied between 31--45 degree, but was consistent with uncertainties. 
Comparing the reduced $\chi^2$ from Table \ref{tab:table2} and Table \ref{tab:table3} shows that the fit obtained by these relativistic spectral models are better than the phenomenological models.

\begin{table*}
\bc
\caption{Broadband X-ray spectral parameters for MAXI J1348--630 from soft state observations fitted with {\tt tbabs} * ({\tt simpl} * {\tt kerrbb} + {\tt relxill}) model. $N_{\rm H}$ is neutral hydrogen column density; $\Gamma$ is asymptotic power-law index; $\rm F_{Sca}$ is scattered fraction of seed photons; $a_*$ is black hole spin parameter; $i$ is inclination angle; {${\dot{m}}$} is effective mass accretion rate of the disk; $\xi$ is ionization of the accretion disk; $A_{\rm Fe}$ is the iron abundance; norm is {\tt relxill} normalization ; $\chi^2/\nu$ is $\chi^2$ per degrees of freedom. The values in the bracket are reduced $\chi^2$. The black hole mass and distance to the source are fixed at 11 M$_\odot$ and 3.39 kpc, respectively.}
{\renewcommand{\arraystretch}{1.3}
	\begin{tabular}{l c c c}
	    \hline
		\hline
Parameter & {Data 1} & {Data 2} & {Data 3} \\\hline
{\tt tbabs} \\
$N_{\rm H}$ [$10^{22}$~cm$^{-2}$] & {0.62 $\substack{+0.01 \\ -0.01}$} & {0.61 $\substack{+0.01 \\ -0.01}$} & {0.59 $\substack{+0.01 \\ -0.01}$}\\\hline
{\tt simpl} \\
$\Gamma$ & {2.01 $\substack{+0.13 \\ -0.09}$} & {2.02 $\substack{+0.11 \\ -0.07}$} & {2.01 $\substack{+0.06 \\ -0.03}$}\\
$\rm F_{Sca}$ [\%] & {2.27 $\substack{+1.70 \\ -1.11}$} & {2.25 $\substack{+1.51 \\ -1.03}$} & {2.32 $\substack{+0.69 \\ -0.73}$} \\
\hline
{\tt kerrbb} \\
$a_*$ & {0.961 $\substack{+0.024 \\ -0.055}$} & {0.994 $\substack{+0.004 \\ -0.034}$} & {0.997 $\substack{+0.001 \\ -0.017}$} \\
$i$ [deg] & {38.2 $\substack{+7.5 \\ -8.4}$} & {32.9 $\substack{+4.9 \\ -1.7}$} & {34.1 $\substack{+4.6 \\ -1.0}$} \\
{${\dot{m}}$} [$\rm 10^{18}~g~s^{-1}$] & {0.52 $\substack{+0.12 \\ -0.21}$} & {0.37 $\substack{+0.09 \\ -0.02}$} & {0.29 $\substack{+0.07 \\ -0.02}$}\\\hline
{\tt relxill} \\
$\log\xi$ [erg cm $s^{-1}$] & {4.13 $\substack{+0.23 \\ -0.42}$} & {4.11 $\substack{+0.22 \\ -0.27}$} & {4.02 $\substack{+0.05 \\ -0.21}$} \\
$A_{\rm Fe}$ & $>$ 2.52 & {5.01 $\substack{+4.49 \\ -2.07}$} & {5.00 $\substack{+2.28 \\ -1.30}$} \\
norm [10$^{-2}$] & {1.024 $\substack{+0.003 \\ -0.002}$} & {1.339 $\substack{+0.004 \\ -0.002}$} & {1.013 $\substack{+0.002 \\ -0.002}$} \\\hline
$\chi^2/\nu$ & 73.2/71 (1.03) & 99.9/71 (1.41)& 100.9/71 (1.42) \\
\hline
\hline
	\end{tabular}}
	\label{tab:table3}
	\ec
\end{table*}
 
\begin{figure}
\centering
   {\includegraphics[width=2.5in,angle=270, trim=0 0 11 0,clip]{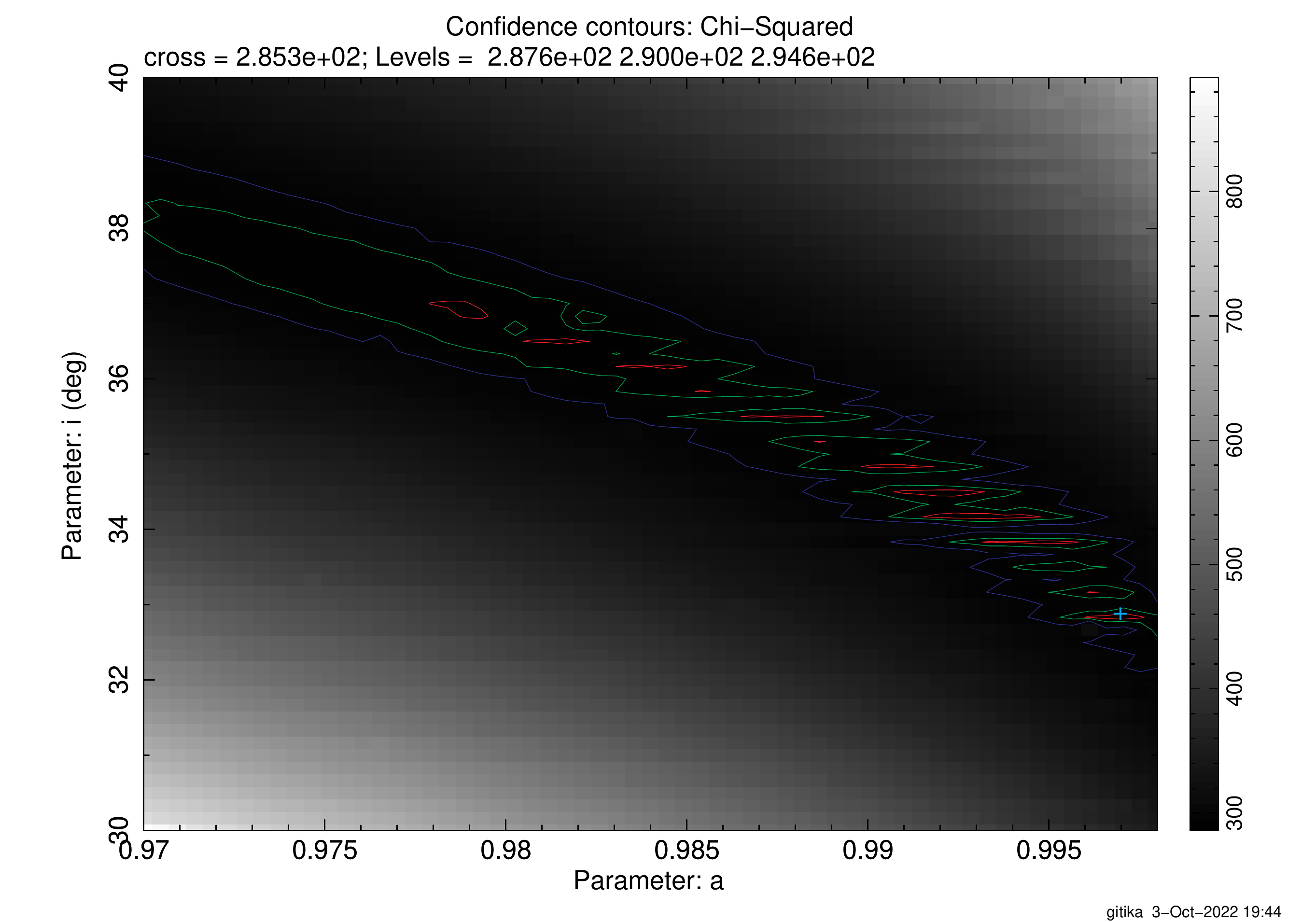}}
\caption{Contour plot of the disc inclination parameter ($i$) and spin parameter ($a_*$) of the source from the joint fit of soft state observations. The 68, 90, and 99 percent contours are shown. The plus mark corresponds to the best-fit local  minimum inferred from the fit.}
  \label{fig:figure3}
  \end{figure}

The spin parameter, the inclination angle and the iron abundance of the source need to be the same for the physical consistency of the fit. For this, we carried out a joint spectral fit for all three observations using {\tt tbabs} * ({\tt simpl} * {\tt kerrbb} + {\tt relxill}). In the joint fit, the spin parameter, the inclination angle and the iron abundance were tied between the spectra, whereas other parameters were allowed to vary between the observations. The best-fit model parameters are listed in Table \ref{tab:table4}. Figure \ref{fig:figure3} shows the confidence contours of disc inclination angle ($i$) and black hole spin parameter ($a_*$).  The joint fit yields the spin parameter to be {0.997 $\substack{+0.001 \\ -0.006}$} however, we see from Figure \ref{fig:figure3} that the parameter value is $>$ 0.97. Thus, the best-fit spin parameter and the inclination angle are $>$ 0.97 and $32.9^{+4.1}_{-0.6}$ degrees, respectively. 

\begin{table*}
\bc
\caption{Broadband X-ray spectral parameters for MAXI J1348--630 from the three soft state observations jointly fitted with {\tt tbabs} * ({\tt simpl} * {\tt kerrbb} + {\tt relxill}) model. The black hole mass and distance to the source are fixed at 11 M$_\odot$ and 3.39 kpc, respectively.}
{\renewcommand{\arraystretch}{1.3}
	\begin{tabular}{l c c c}
	    \hline
		\hline
Parameter & {Data 1} & {Data 2} & {Data 3} \\\hline
{\tt tbabs} \\
$N_{\rm H}$ [$10^{22}$~cm$^{-2}$] & {0.61 $\substack{+0.01 \\ -0.01}$} & {0.61 $\substack{+0.01 \\ -0.01}$} & {0.61  $\substack{+0.01 \\ -0.01}$}\\\hline
{\tt simpl} \\
$\Gamma$ & {2.00$\substack{+0.09 \\ -0.06}$} & {1.99 $\substack{+0.08 \\ -0.06}$} & {2.06 $\substack{+0.05 \\ -0.09}$}\\
$\rm F_{Sca}$ [\%] & {2.15 $\substack{+1.34 \\ -0.83}$} & {2.02 $\substack{+1.32 \\ -0.89}$} & {2.89 $\substack{+0.94 \\ -1.12}$} \\
\hline
{\tt kerrbb} \\
$a_*$ & {0.997 $\substack{+0.001 \\ -0.006}$} & {0.997 $\substack{+0.001 \\ -0.006}$} & {0.997 $\substack{+0.001 \\ -0.006}$} \\
$i$ [deg] & {32.9 $\substack{+4.1 \\ -0.6}$} & {32.9 $\substack{+4.1 \\ -0.6}$} & {32.9 $\substack{+4.1 \\ -0.6}$} \\
{${\dot{m}}$} [$\rm 10^{18}~g~s^{-1}$] & {0.38 $\substack{+0.05 \\ -0.02}$} & {0.35 $\substack{+0.13 \\ -0.11}$} & {0.31 $\substack{+0.04 \\ -0.01}$}\\\hline
{\tt relxill} \\
$\log\xi$ [erg cm $s^{-1}$] & {4.11 $\substack{+0.07 \\ -0.07}$} & {4.11 $\substack{+0.16 \\ -0.06}$} & {4.00 $\substack{+0.05 \\ -0.19}$} \\
$A_{\rm Fe}$ & {5.01 $\substack{+2.57 \\ -2.33}$} & 5.01 {$\substack{+2.57 \\ -2.33}$} & 5.01 {$\substack{+2.57 \\ -2.33}$} \\
norm [10$^{-2}$] & {1.178 $\substack{+0.252 \\ -0.237}$} & {1.318 $\substack{+0.211 \\ -0.212}$} & {1.087 $\substack{+0.184 \\ -0.172}$} \\\hline
$\chi^2/\nu$ & & 285.7/219 (1.31) & \\
\hline
\hline
	\end{tabular}}
	\label{tab:table4}
	\ec
\end{table*}
 
\subsection{Hard State Observations}

The 0.3--8 keV SXT and 3--80 keV LAXPC light curves from hard state observations (Data4 and Data5) are depicted in Figure \ref{fig:figure5}. The spectral analysis of the hard state observations was performed as mentioned in \S \ref{sec:3.1}. However, we replaced the {\tt kerrbb} component from the model combination described in \S \ref{sec:3.1} with the relativistic disc model {\tt kerrd} (\citealt{2003ApJ...597..780E}) to allow for the possibility that the inner disc radius may not be at ISCO. {\tt Kerrd} accommodates the mass, disc inclination angle, accretion rate as well as the inner radius of the disc. The spectral hardening factor, the normalization and the outer radius of the disc were frozen at 1.7, 1, and $10^{4}$ r\textsubscript{g}, respectively. In addition, we fixed the inclination angle, the spin parameter and the iron abundance to the values obtained from the joint fit of all the soft state observations (see Table \ref{tab:table4}). The inner disc radius of {\tt kerrd} was tied to inner disc radius in {\tt relxill} after multiplying by a factor of 1.278, since for {\tt relxill} it is in units of the radius of marginal stability, whereas for {\tt kerrd} the radius is calculated in r\textsubscript{g}. Note that for $a_*$ = 0.997, the radius of marginal stability is 1.278 r$_{g}$.

\begin{figure*}
\centering
	\includegraphics[width=2.9in,angle=0,trim=0 0 0 0,clip]{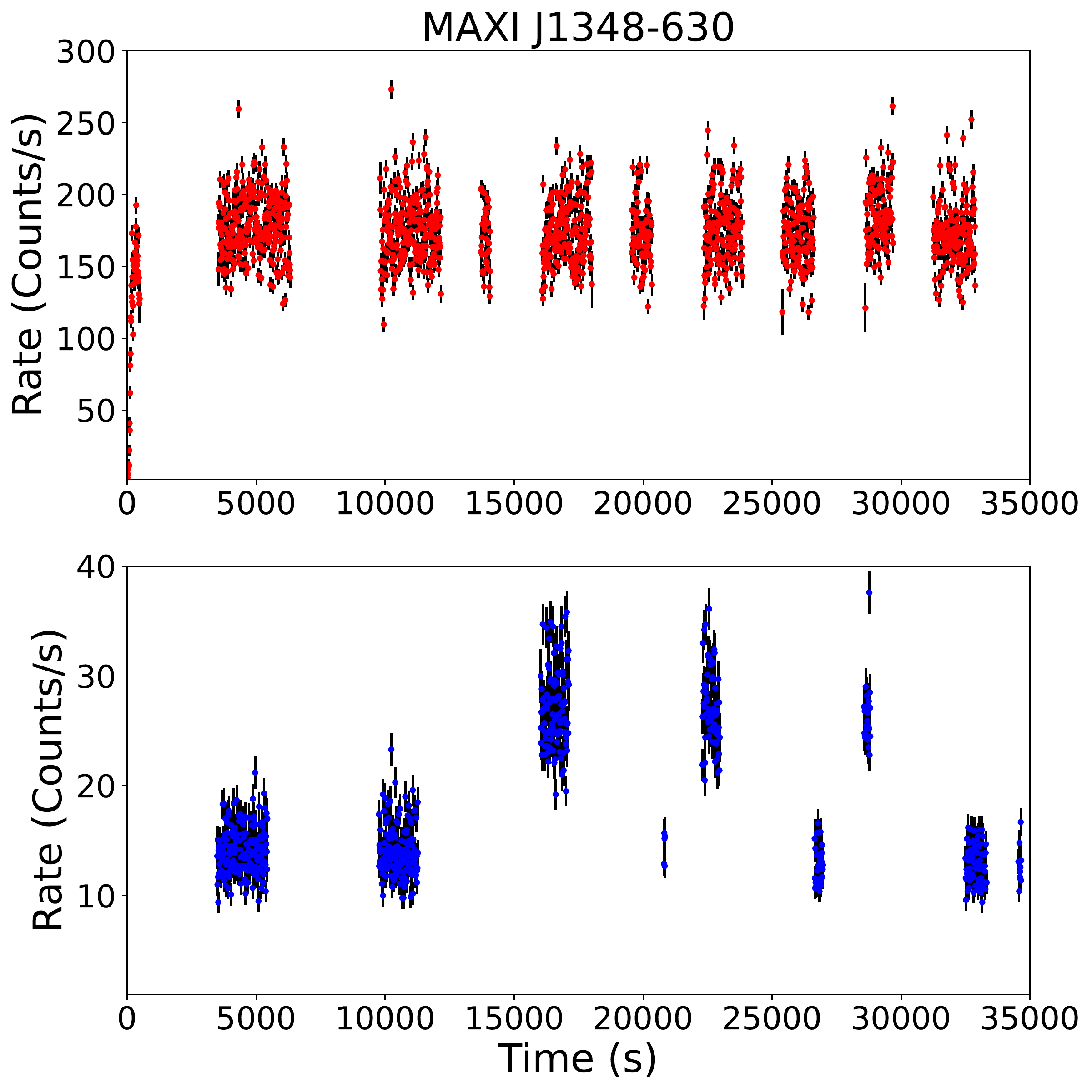}
	\includegraphics[width=2.9in,angle=0,trim=0 0 0 0,clip]{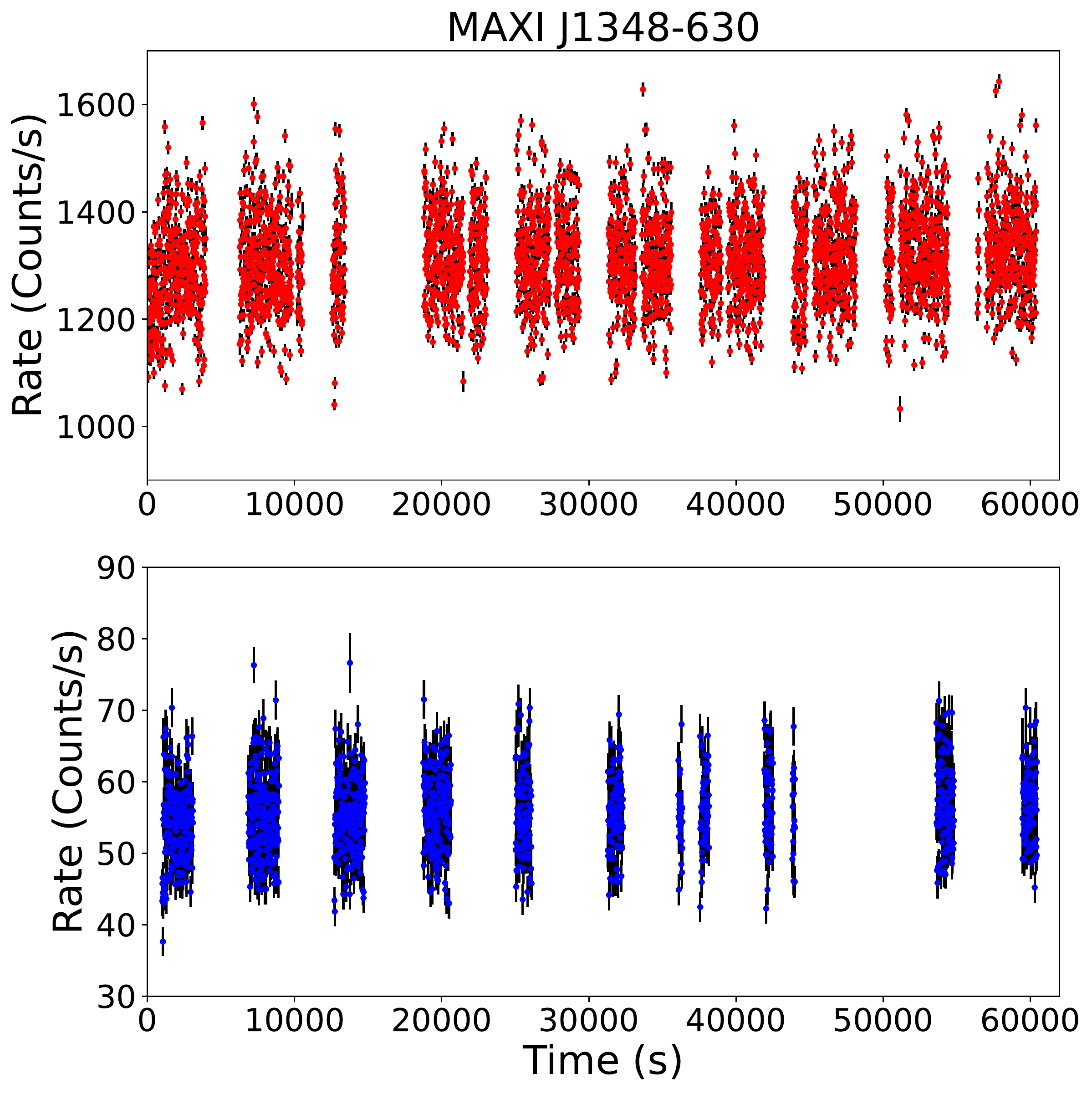}
    \caption{The light curves from {\it AstroSat} observations in the hard state: Data4 (left) and Data5 (right). In each plot, the top and bottom panels represent the 3--80 keV light curve from LAXPC20 binned with 10-s and the 0.3--8 keV light curve from SXT, respectively. The reference time for Data 4 is MJD 58611 and for Data 5 is MJD 58648.}
    \label{fig:figure4}
\end{figure*}

The unfolded spectra and residuals for the two observations are depicted in Figure \ref{fig:figure5} and the best-fit model parameters are listed in Table \ref{tab:table5}. The absorption column density  yields $\sim 5 \times 10^{21}~\rm cm^{-2}$. The photon index is $<$ 1.52 for both observations. The mass accretion rate decreased by an order of magnitude compared to the soft state observations and is $\sim 1-2 \times \rm 10^{16}~g~s^{-1}$. The best fit values for R\textsubscript{in} for Data4 is $\sim 4.9 $r$_{g}$ and hence the disc seems to be truncated. However, for Data5 the radius comes out to be close to ISCO.  The $\chi^{2}$ distributions as a function of R\textsubscript{in} for Data4 and 5 are shown in Fig. \ref{fig:figure6}. These results are further discussed in the next section.

\begin{figure*}
\centering
{\includegraphics[width=2.1in,angle=270,trim=0 0 0 0,clip]{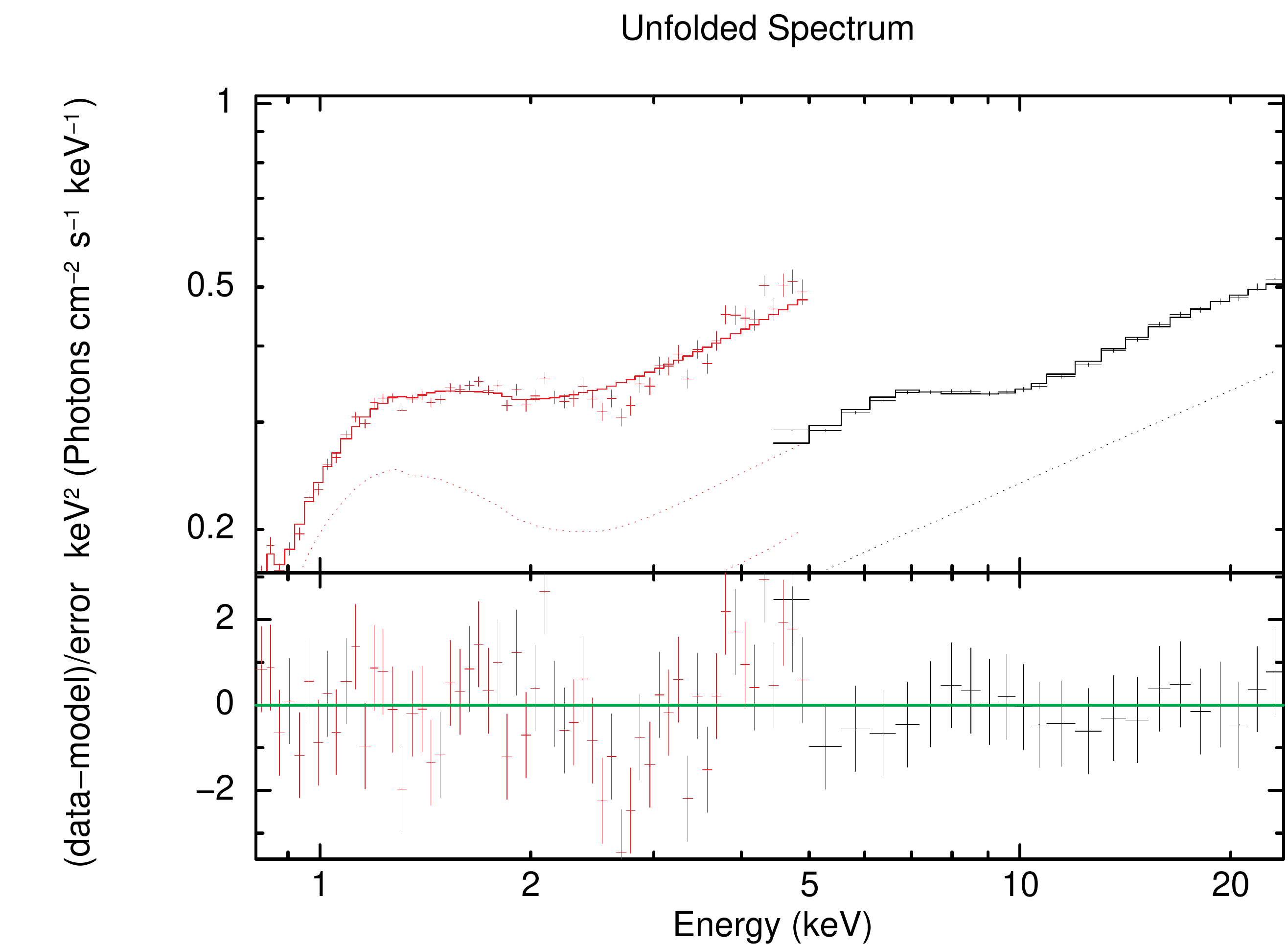}}	{\includegraphics[width=2.1in,angle=270,trim=0 0 0 0,clip]{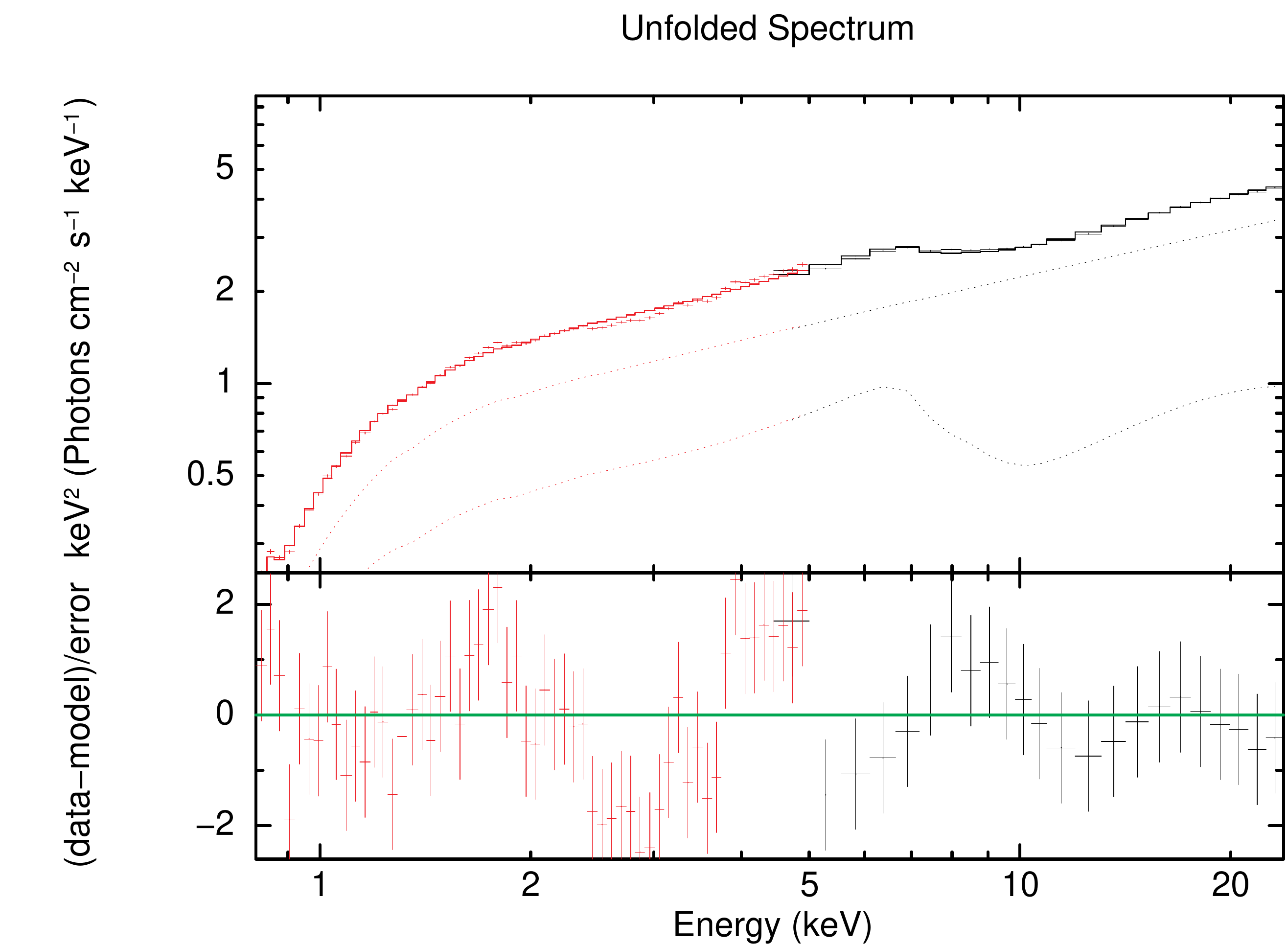}}
    \caption{The 0.8–25 keV broadband X-ray  unfolded spectra of MAXI J1348--630 for Data4 (left panel) and Data5 (right panel). The black and red data points represent LAXPC20 and SXT data, respectively. Spectra are fitted with {\tt tbabs} * ({\tt simpl} * {\tt kerrd} + {\tt relxill}).}
    \label{fig:figure5}
\end{figure*}

\begin{figure*}
\centering
	{\includegraphics[width=3.5in,angle=0,trim=0 0 0 0,clip]{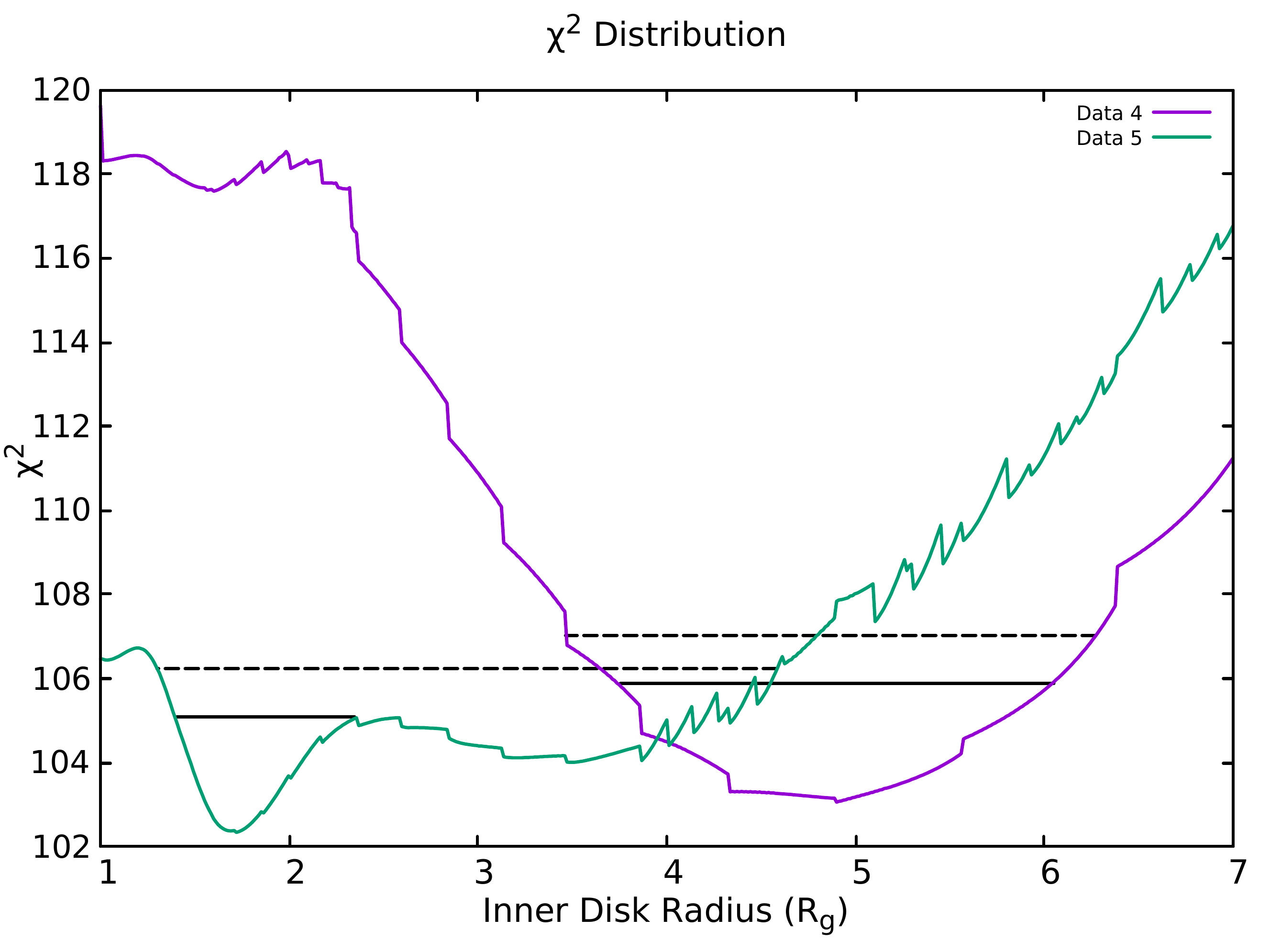}}
    \caption{ The $\chi^{2}$ distributions as a function of R\textsubscript{in} when the 0.8–25 keV broadband X-ray spectrum of MAXI J1348--630 fitted with {\tt tbabs} * ({\tt simpl} * {\tt kerrd} + {\tt relxill}). The purple and green curves represent Data4 and Data5, respectively. The non-dashed and dashed lines show 3-$\sigma$ and 4-$\sigma$ levels.}
    \label{fig:figure6}
\end{figure*}

\begin{table*}
\bc
\caption{Broadband X-ray spectral parameters for MAXI J1348--630 from hard state observations fitted with {\tt tbabs} * ({\tt simpl} * {\tt kerrd} + {\tt relxill}) model. The black hole mass and distance to the source are fixed at 11 M$_\odot$ and 3.39 kpc, respectively.}
{\renewcommand{\arraystretch}{1.3}
	\begin{tabular}{l c c c}
	    \hline
		\hline
Parameter & {Data 4} & {Data 5} \\\hline
{\tt tbabs} \\
$N_{\rm H}$ [$10^{22}$~cm$^{-2}$] & {0.44 $\substack{+0.03 \\ -0.04}$} & {0.51 $\substack{+0.01 \\ -0.01}$}\\\hline
{\tt simpl} \\
$\rm F_{Sca}$ [\%] &  {10.4 $\substack{+2.2 \\ -1.4}$} &  {51.8 $\substack{+2.4 \\ -2.4}$} \\
\hline
{\tt kerrd} \\
{${\dot{m}}$} [$\rm 10^{17}~g~s^{-1}$] & {0.12 $\substack{+0.04 \\ -0.03}$} & {0.19 $\substack{+0.07 \\ -0.01}$}\\\hline
{\tt relxill} \\
R$_{in}$ [r$_{g}$] &  {4.9 $\substack{+1.2 \\ -1.1}$} &  {1.7 $\substack{+2.7 \\ -0.3}$} \\
$\Gamma$ & $<$ 1.52 & $<$ 1.52 \\
$\log\xi$ [erg cm $s^{-1}$] &  {3.8 $\substack{+0.3 \\ -0.1}$} &  {3.4 $\substack{+0.1 \\ -0.1}$}\\
norm [10$^{-3}$] &  {0.44 $\substack{+0.05 \\ -0.06}$} &  {4.14 $\substack{+0.11 \\ -0.61}$} \\\hline
$\chi^2/\nu$ & 103.2/73 (1.41) & 102.4/73 (1.40)  \\
\hline
\hline
	\end{tabular}}
	\label{tab:table5}
	\ec
\end{table*}

\section{Discussion}
\label{sec:4}

In this work, we performed detailed broadband spectral analysis of the transient BHXRB MAXI J1348--630 using {\it AstroSat} observations. Using a general relativistic accretion disc and iron line emission along with a thermal Comptonization component, we described the soft state spectra of the source and estimated important system parameters. The joint analysis of soft state observations suggests that the source hosts a rapidly spinning black hole with  $a_*$ $>$ 0.97. The disc inclination angle, constrained at $i = 32.9 \substack{+4.1\\-0.6}$ degrees, is consistent with the values reported by \citet{2021MNRAS.508..475C}. 

The three soft state observations are spectrally similar without showing any significant difference in the parameter values (Table \ref{tab:table4}). On the other hand, \citet{Jithesh_2021} shows that there are significant differences in the power density spectra of the three observations in the LAXPC (3--15 keV) band. In fact, different segments of the Data2 have significantly different variability amplitude. As expected for the soft state, the variability in the {\it NICER} (0.5--12 keV) band is significantly smaller than the higher energy LAXPC band. Thus the disc is relatively stable and the variability  that seems to independent of the disc spectral parameters can be attributed to different temporal properties of the corona. \\

To better understand the disc truncation in the hard state, we used the relativistic disc model {\tt kerrd} for the hard state observations. For the low hard state we obtain a more truncated accretion disk than that for the bright hard state observation. There, the inner radii was estimated to be closer to the black hole. The disc accretion rate was lower by a factor of ten than that of the soft state. \citet{2021MNRAS.508..475C} and \citet{2022MNRAS.513.4869K} have also estimated accretion rates for soft and hard state observations with a difference of a factor of ten. \citet{2022ApJ...927..210Z} too reported an unusual behavior where the inner radius, during the rising hard state, was found to be smaller than in the soft state.
In addition, for the bright hard state observation, the estimated unabsorbed bolometric luminosity was lower by a factor of three than the soft state, while for the faint hard state observation it was lower, by more than a factor of ten.\\
An inner disc radii close to the black hole has been reported for the hard spectral state of three other black hole systems: Swift J1753.5-0127, GRS 1739--278 and Cygnus X--1. \citet{Miller_2006} analysed Swift J1753.5--0127 using {\it XMM-Newton} and {\it RXTE} observations, 
and reported that the disc remained close to the ISCO in the low hard state, during both the rising and decaying  phases of the outburst. In the case of GRS 1739--278, {\it NuSTAR} observations were conducted during the peak of rising low hard state which revealed a broad, skewed iron line and disc reflection spectrum \citep{Miller_2015}. Modelling the {\it NuSTAR} spectrum with different relativistic reflection models suggested that the accretion disc extends close to the black hole and the estimated inner radius was $\sim 5$ r\textsubscript{g}. In the bright hard state of Cygnus X--1, during the 2014 observations, the disc did not truncate at a large distance from the compact object, instead the disc remained close to the ISCO with a inner radius of R\textsubscript{in} $<$ 1.7 r\textsubscript{ISCO} \citep{Parker_2015}. However, \citet{10.1093/mnras/stx2283} reported the presence of a substantially truncated accretion disc at $\sim$ 13--20 r\textsubscript{g} for Cygnus X--1 by using a different spectral model. In the case of MAXI J1348--630, we find that the disc accretion disc extends very close to the black hole with an inner radius of $<$ 2 r\textsubscript{g} for the bright hard state observation. Nevertheless, we emphasise that the disc component is weak for the hard state and hence inner radii measurements maybe limited by spectral resolution of the instruments and spectral complexity (such as the addition of a low temperature Comptonization component) can affect the estimation (\citealt{10.1093/mnras/stx2283}).\\

MAXI J1348-630 has been observed exhibiting a type-B QPO whose transitions and plausible origins have been studied by several authors. \cite{2021MNRAS.505.3823Z} and \cite{https://doi.org/10.48550/arxiv.2208.07066} presented a spectral-timing analysis of the fast appearance/disappearance of a type-B QPO observed in NICER and \textit{Insight}-HXMT observations of MAXI J1348-−630 respectively. \textit{Insight}–HMXT observed a type-C/B transition that occurs within a very short time scale. \cite{https://doi.org/10.48550/arxiv.2208.07066} summarised that the variation amplitude (rms) of QPOs increases with energy below 10 keV and remains constant above from 10 keV to 100 keV. From their results, they associated the appearance and disappearance of type-B QPO with a change in the flux of the Comptonization component, the inner disk radius, and the corona height. They suggest that type-B QPOs probably originate from the precession of the weak-jet in a titled disk-jet structure located relatively close to the compact object. \cite{2021MNRAS.505.3823Z} found that the type-B QPO is associated with a redistribution of accretion power between the disk and Comptonised emission. The energy-dependent fractional rms and phase lags of the type-B QPO here have been explained by a two-component Comptonisation model (\citealt{2020MNRAS.492.1399K}) by \cite{2021MNRAS.501.3173G} They gave consideration  to the possibility that the variability spectra of the type-B QPO arise from Comptonisation occurring in two (a small and a large region) different, but physically connected, regions located in the vicinity of the compact object with a black-body source of soft photons. They found that the large component dominates the variability below $\sim$ 2 keV, whereas above that energy the small component prevails in harder energies. Their best-fitting time-dependent Comptonising model consisted of a small region, with rather low feedback, and a large region, with high feedback. \cite{Bellavita_2022} followed the same process using the same model but here the soft photon source is the accretion disk modelled with {\tt diskbb}. Their model {\tt vkdualdk} fits the data very well. Their best-fitting model yields a small Comptonisation region with very high feedback and a large Comptonisation region with a relatively low feedback fraction.\\

In our study, we reported a difference of more than a factor of six between the luminosities of the bright hard state and the faint one. Furthermore, the accretion rate in the disc is inferred to be lower, while the inner disc radius is larger for the faint state compared to the bright one. \citet{Jithesh_2021} reported the r.m.s and time-lag as a function of energy for the hard state observations and modelled them using a single zone Comptonization model (\citealt{10.1093/mnras/stz930}). Variations of the inner disc temperature ($\delta kT_{in}$) and the coronal temperature ($\delta kT_{e}$) occur with a time difference ($\tau_d$) between them which gives rise to the observed energy-dependent r.m.s and time-lag. \citet{Jithesh_2021} observed that $\tau_d$ is positive ($\sim 70$ ms) for the bright state, but it is negative $\tau_d \sim -90$ ms, for the faint one when variations at $\sim 1$ Hz are considered. This reversal in the sign of the time lag implies that the temporal causality (i.e. whether the disc temperature varies before the coronal temperature or vice versa) has changed between the two observations. It is tempting to connect this with the larger inner radius for the faint state suggested in the spectral fitting of this work, however, it is not clear how such a change in geometry will cause the causality to be reversed. Moreover, as mentioned earlier, the estimation of the inner radii during the hard state is subject to uncertainties in the spectral resolution of the instruments and the complexity of the spectral models used to fit the data. Nevertheless, this study underlines the potential of conducting spectral and timing analysis of a black hole system as it evolves through different states using simultaneous observations from different instruments.

\section*{Acknowledgements}

We thank the anonymous referee for the constructive comments and suggestions that improved the manuscript. GM acknowledges the support from the China Scholarship Council (CSC), Grant No. 2020GXZ016647, and IUCAA Visitors Programme. GM would also like to thank Dr. Yash Bhargava for the insightful discussions. VJ would like to acknowledge the Centre for Research, CHRIST (Deemed to be University), for the financial support in the form of a Seed Money Grant. The research is based on the observations acquired by the {\it AstroSat} mission of the Indian Space Research Organization (ISRO), archived at the Indian Space Science Data Centre (ISSDC). This study has made use of the data whence the LAXPC and SXT instruments. We show courtesy to the LAXPC Payload Operation Center (POC) and the SXT POC at TIFR, Mumbai for dispensing the data from the ISSDC data archive and the imperative software tools. This research has used the data and software provided by the High Energy Astrophysics Science Archive Research Center (HEASARC), which is a service of the Astrophysics Science Division at NASA/GSFC.
\section*{DATA AVAILABILITY}
The data used in this work is available in the ISRO’s Science Data Archive for {\it AstroSat} Mission ({\href{https://
astrobrowse.issdc.gov.in/astro$\_$archive/archive/Home.jsp}{https://
astrobrowse.issdc.gov.in/astro$\_$archive/archive/Home.jsp)}} . 

\bibliographystyle{raa}
\bibliography{J1348-raa-draft}

\label{lastpage}

\end{document}